\documentclass[prb,preprint]{revtex4-1} 
\usepackage{amsmath}  % needed for \tfrac, \bmatrix, etc.
\usepackage{amsfonts} % needed for bold Greek, Fraktur, and blackboard bold
\usepackage{graphicx} % needed for figures
\usepackage{hyperref} % needed for adding hyperlink

\begin{document}

% Be sure to use the \title, \author, \affiliation, and \abstract macros
% to format your title page.  Don't use lower-level macros to  manually
% adjust the fonts and centering.

\title{Everyday Radio Telescope}
% In a long title you can use \\ to force a line break at a certain location.

% If there were a second author at the same address, we would put another 
% \author{} statement here.  Don't combine multiple authors in a single
% \author statement.
%\affiliation{Department of Physics, Weber State University, Ogden, UT 84408-2508}
% Please provide a full mailing address here.

\author{Pranshu Mandal}
\email{prnshmandal@iisertvm.ac.in}
\affiliation{Indian Institute of Science Education and Research Thiruvananthapuram, CET Campus, Trivandrum 695016}

\author{Devansh Agarwal}
\email{devansh@iisertvm.ac.in}
\affiliation{Indian Institute of Science Education and Research Thiruvananthapuram, CET Campus, Trivandrum 695016}

\author{Pratik Kumar}
\email{pratik13@iisertvm.ac.in}
\affiliation{Indian Institute of Science Education and Research Thiruvananthapuram, CET Campus, Trivandrum 695016}

\author{Anjali Yelikar}
%\email{kanchan2613@iisertvm.ac.in}
\affiliation{Indian Institute of Science Education and Research Thiruvananthapuram, CET Campus, Trivandrum 695016}

\author{Kanchan Soni}
\affiliation{Indian Institute of Science Education and Research Thiruvananthapuram, CET Campus, Trivandrum 695016}
\author{Vineeth Krishna T}
\affiliation{Indian Institute of Science Education and Research Thiruvananthapuram, CET Campus, Trivandrum 695016}

% See the REVTeX documentation for more examples of author and affiliation lists.

\date{\today}

\begin{abstract}
We have developed an affordable, portable college level radio telescope for amateur radio astronomy which can be used to provide hands-on experience with the fundamentals of a radio telescope and an insight into the realm of radio astronomy. With our set-up one can measure brightness temperature and flux of the Sun at 11.2 GHz and calculate the beam width of the antenna. The set-up uses commercially available satellite television receiving system and parabolic dish antenna. We report the detection of point sources like Saturn and extended sources like the galactic arm of the Milky way. We have also developed python pipeline, which are available for free download, for data acquisition and visualization.
\end{abstract}
% AJP requires an abstract for all regular article submissions.
% Abstracts are optional for submissions to the "Notes and Discussions" section.

\maketitle % title page is now complete

%%%%%%%%%%%%%%%%%%%%%%%%%%%%%%%%%%%%%%%%%%%%%%%%%%%%%%%%%%%%%%%%%%%%%%%%%%%%%%%%%%%%%%%%%%%%%%%%%%%%%%%%%%%%%%%%%%%%%%%%%%%%%
\section{Introduction} % Section titles are automatically converted to all-caps.
Radio waves from astronomical sources have low intensities, and due to their long wavelengths, large radio telescopes are needed to gather signal and achieve resolutions comparable to optical telescopes. Building large telescopes is mechanically and economically challenging. Several schools, colleges and community astronomy clubs have optical telescopes, however, radio telescopes are not as famous due to their cost, the size and complexity of the data. We have developed a small, affordable radio telescope and data analysis tools to do amateur radio astronomy at such levels.\\

For demonstrating the usefulness of such a set-up, we have measured the brightness temperature of the sun. We present how this can be used to measure the flux density of the sun and correlate them with presently available data. Modifications in the observation techniques, which improved the sensitivity and enabled us to detect point sources such as Saturn and extended sources such as the galactic arm, is also discussed here.
%%%%%%%%%%%%%%%%%%%%%%%%%%%%%%%%%%%%%%%%%%%%%%%%%%%%%%%%%%%%%%%%%%%%%%%%%%%%%%%%%%%%%%%%%%%%%%%%%%%%%%%%%%%%%%%%%%%%%%%%%%%%%
\section{Aims and Set-up}
The prime motivation behind this project was to assemble and operationalize a low-cost radio telescope for amateur radio astronomy, study the properties of the telescope components and finally establish an undergraduate level experiment in radio astronomy. One that teaches the rudiments of this very well established technology by providing the necessary experiments.\\

The setup has been proposed and assembled previously from NCRA(ASRT)\cite{ASRT}, MIT(VSRT)\cite{VSRT}, NRAO(IBT)\cite{NRAO}, which was initially motivated by the works of Cleary et. al(1993) \cite{Cleary}. This provided the groundwork to build on. A block diagram of our assembled setup is shown in \autoref{setupcircuit}. The key components of the set up are:
\begin{figure}[h]
\centering
\includegraphics[scale=0.6]{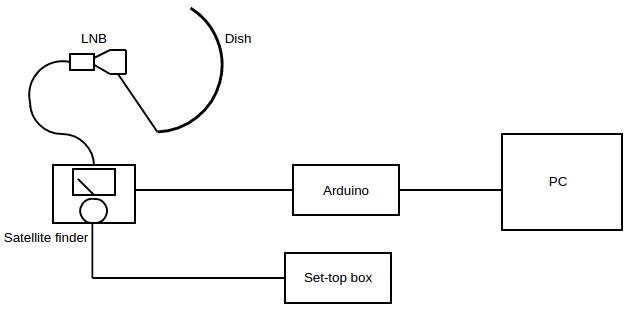}
\caption{Block diagram of the setup.}
\label{setupcircuit}
\end{figure}
\begin{enumerate}
\item Satellite Antenna: Satellite dish antenna consisting of a 68 cm parabolic dish reflector and an LNB (Low Noise Block). The LNB receives the radio signal from the satellite reflected by the dish and amplifies it. A set-top box powers the satellite finder and the LNB.
\item Satellite Finder: A satellite signal meter is commonly used for orienting satellite dishes towards geostationary satellites. We have used a commercially available analog satellite finder (\textbf{GC SF-02}). Such a device acts as a square law detector and is used to read directly the intensity.
\item Arduino Uno: Arduino is a microcontroller board with a 10 bit Analog to Digital Converter for us. Using this board, we digitize the intensity from the satellite finder at 10Hz sampling rate. 
\end{enumerate}
The python pipeline for data acquisition and visualization can be found online.\cite{pipeline}
%%%%%%%%%%%%%%%%%%%%%%%%%%%%%%%%%%%%%%%%%%%%%%%%%%%%%%%%%%%%%%%%%%%%%%%%%%%%%%%%%%%%%%%%%%%%%%%%%%%%%%%%%%%%%%%%%%%%%%%%%%%%%
\section{Analysis}
\label{analysis}
Geostationary satellites emit polarized radiation, after pointing in the direction of satellites, the LNB was rotated to align the monopole inside in the direction of polarized radiation maximizing the detected signal intensity. Pointing was done manually in all the cases, by firstly looking at the source Altitude and Azimuth in the local sky at that time, using Stellarium and then rotating the dish about the two axes individually to point at the source location. The telescope was calibrated with geostationary satellites and power per unit voltage was calculated. For the temperature calculation, power density was calculated given aperture efficiency followed by the flux calculation. By assuming the sun's radiation to be thermal, the temperature was calculated following Rayleigh-Jeans law.\cite{stix} See \autoref{appendix1} for detailed calculations.\\

\begin{figure}[h]
\centering
\includegraphics[scale=0.65]{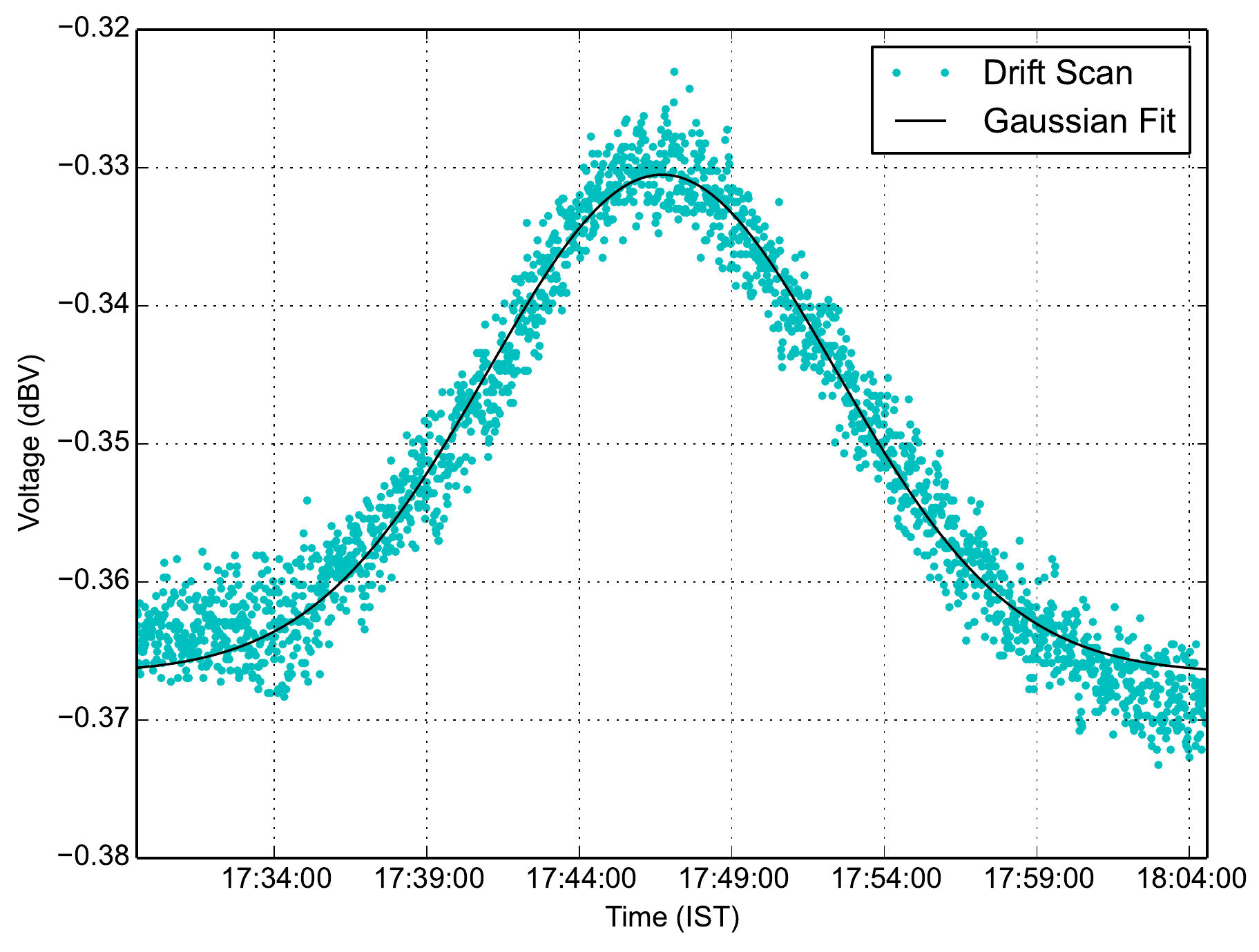}
\caption{Drift scan of the Sun. Cyan dots is the data and black line is a Gaussian fit of the data.}
\label{driftscan}
\end{figure}

For beam-width calculation, the voltage data (in dBV) was fitted with a Gaussian ($f\left(t\right) = a e^{- { \frac{(t-b)^2 }{ 2 c^2} } }$) using python and Full Width at Half Maxima (FWHM) was calculated using the variance ($\mathrm{FWHM} = 2 \sqrt{2 \ln 2}\ c$). \autoref{table3} provides the obtained values. \autoref{driftscan}, shows the drift scan of the sun. A Gaussian was fit to the data and beam width of our telescope was characterized to be $ \sim 3.5^{\circ} $ with less than 1\% error. These are in agreement with standard online beam width calculators\cite{beamwidth} which estimate beam width of our system to be $ \sim 3.4^{\circ}$.
\begin{table}[!h]
\centering
\begin{tabular}{|c|c|}
\hline
Epoch(MJD) &Beam width(degree) \\
\hline
57111&3.357 $\pm$ 0.021\\
\hline
57108&3.501 $\pm$ 0.007\\
\hline
\end{tabular}
\caption{Beam width calculated from graphs}
\label{table3}
\end{table}
For the noise analysis, the dish was pointed at the cold sky. By cold sky, we refer to the continuous radiation coming from the sky in absence of a source in the field of view of the telescope. This signal is predominantly due to the presence of Cosmic Microwave Background. A Fourier fit was performed, and residuals were termed as noise and taken forward for FFT in MATLAB. To evaluate the system generated noise, the LNB was shielded with thick aluminium foils to block any external radiation from reaching the system, and the same procedure was repeated.\\

\begin{figure}
\centering
\includegraphics[scale=0.75]{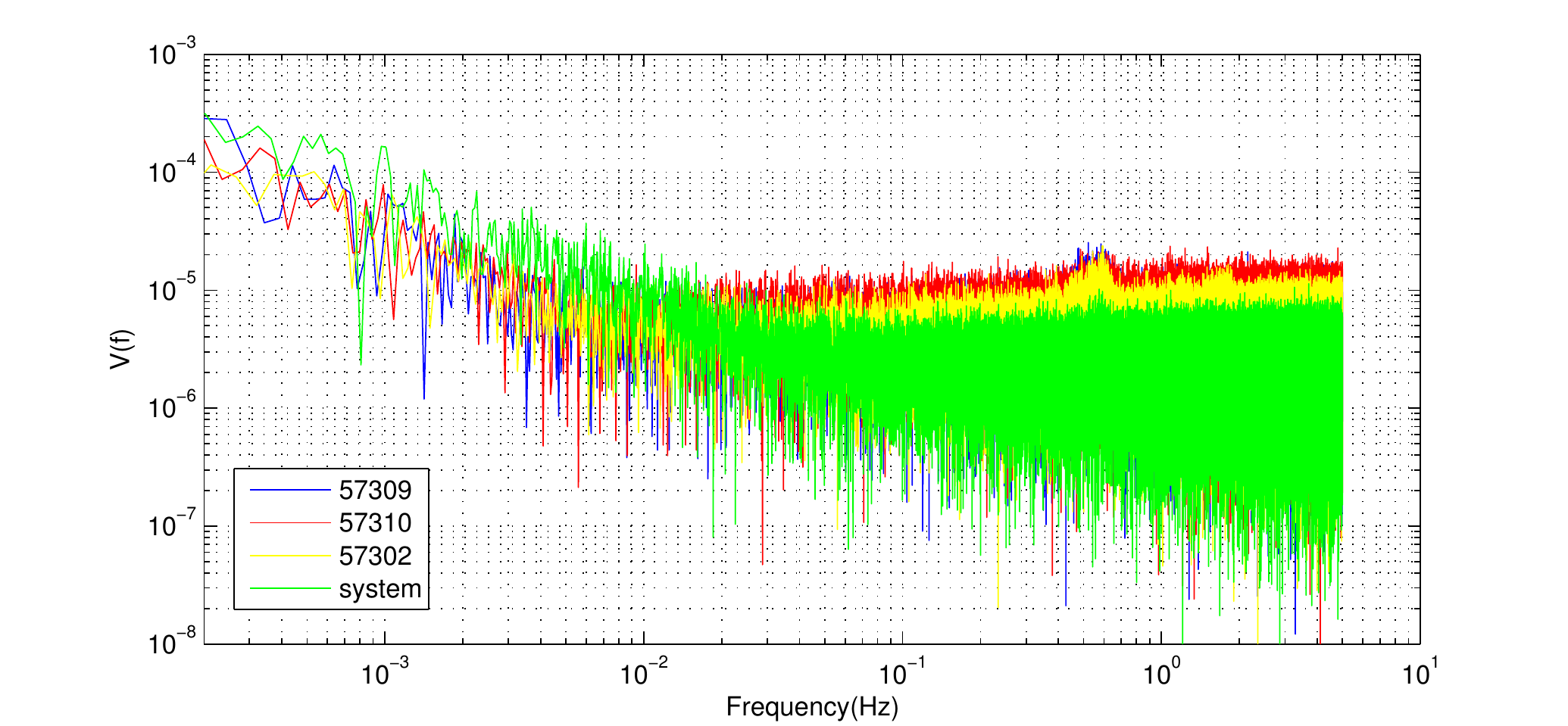}
\caption{FFT of the daytime data. Different colour shows different epochs}
\label{fig:my_label1}
\includegraphics[scale=0.75]{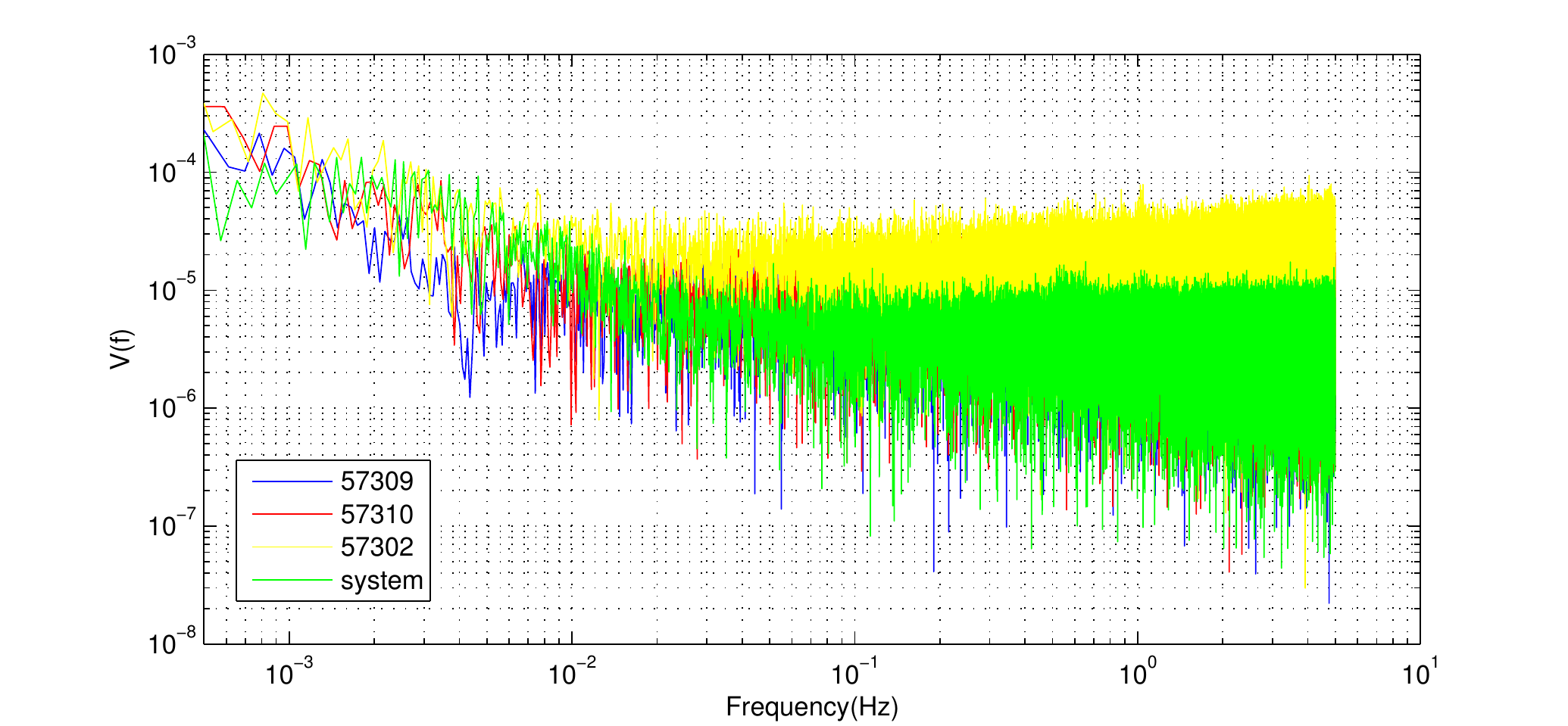}
\caption{FFT of the night data. Different colour shows different epochs}
\label{fig:my_label2}
\end{figure}

The noise in the signal (from the cold sky) is dominated by the system generated noise. The features in the FFT spectra (like a bump near $\sim$ 0.5 Hz) of system generated noise gets carried forward in the FFT spectra of the cold sky. The FFT spectra also revealed a difference in the power levels of the signal and the system generated noise. The signal had a higher  noise power, which was likely to be observed.
A difference in power levels of FFT spectrum for various days during day and night time was also noticed. This difference could be ascribed to temperature variations during the period of observation. \autoref{fig:my_label1} and \autoref{fig:my_label2} show the FFT spectrum during night and day time respectively.
%%%%%%%%%%%%%%%%%%%%%%%%%%%%%%%%%%%%%%%%%%%%%%%%%%%%%%%%%%%%%%%%%%%%%%%%%%%%%%%%%%%%%%%%%%%%%%%%%%%%%%%%%%%%%%%%%%%%%%%%%%%%%
\section{Results}
\subsection{Solar Brightness Temperature}
The flux received from the Sun was calibrated against the standardized data from various geostationary satellites, NSS 6, SES 7 and INSAT 4A. The flux received from these satellites and their corresponding voltages used in the calculation are shown in \autoref{table1} and \autoref{table2}. The brightness temperature of the sun was calculated to near 10,000 K with a maximum error of 3.23 \%, which is slightly lower than the values reported by Zirin et. al (1991)\cite{Zirin}. Details of calculation are given in appendix \ref{appendix1}. 
\begin{table}[h]
\centering
\begin{tabular}{|c|c|c|}
\hline
Name Of Geo-Sat & Altitude (Km) & EIRP (dBW)\\
\hline
NSS6&36237&50\\
\hline
SES7&36934&53\\
\hline
INSAT 4A&35916&52\\
%\hline
%ST 2 &73.6&36012&129.6&51\\
\hline
\end{tabular}
\caption{Table of a few Geostationary satellites and their corresponding Equivalent isotropically radiated power (EIRP) used for calibrating the setup }
\label{table1}
\end{table}
\begin{table}[h]
\begin{tabular}{|c|c|c|c|c|c|c|c|}
\hline
Epoch(MJD)&$ \Delta $ Vsat(V)&$ \Delta $ Vsun (V)&$ \sigma $ Cold Sky(V)&$ \sigma $ Sun(V)&$ \sigma $ Sat(V)&Temp(K)&$ \delta $ T(K)\\
\hline
57099&0.0980&0.0521&0.0009, 0.0009&0.0008&0.0016&10078&295\\
\hline
57115&0.0922&0.0493&0.0009, 0.0009&0.0011&0.0011&10136&324\\
\hline
\end{tabular}
\caption{Table of temperature calculated from same day calibration and scan data}
\label{table2}
\end{table}
\subsection{Saturn and Milky Way Detection}
\begin{figure}[h]
\centering
\includegraphics[scale=0.65]{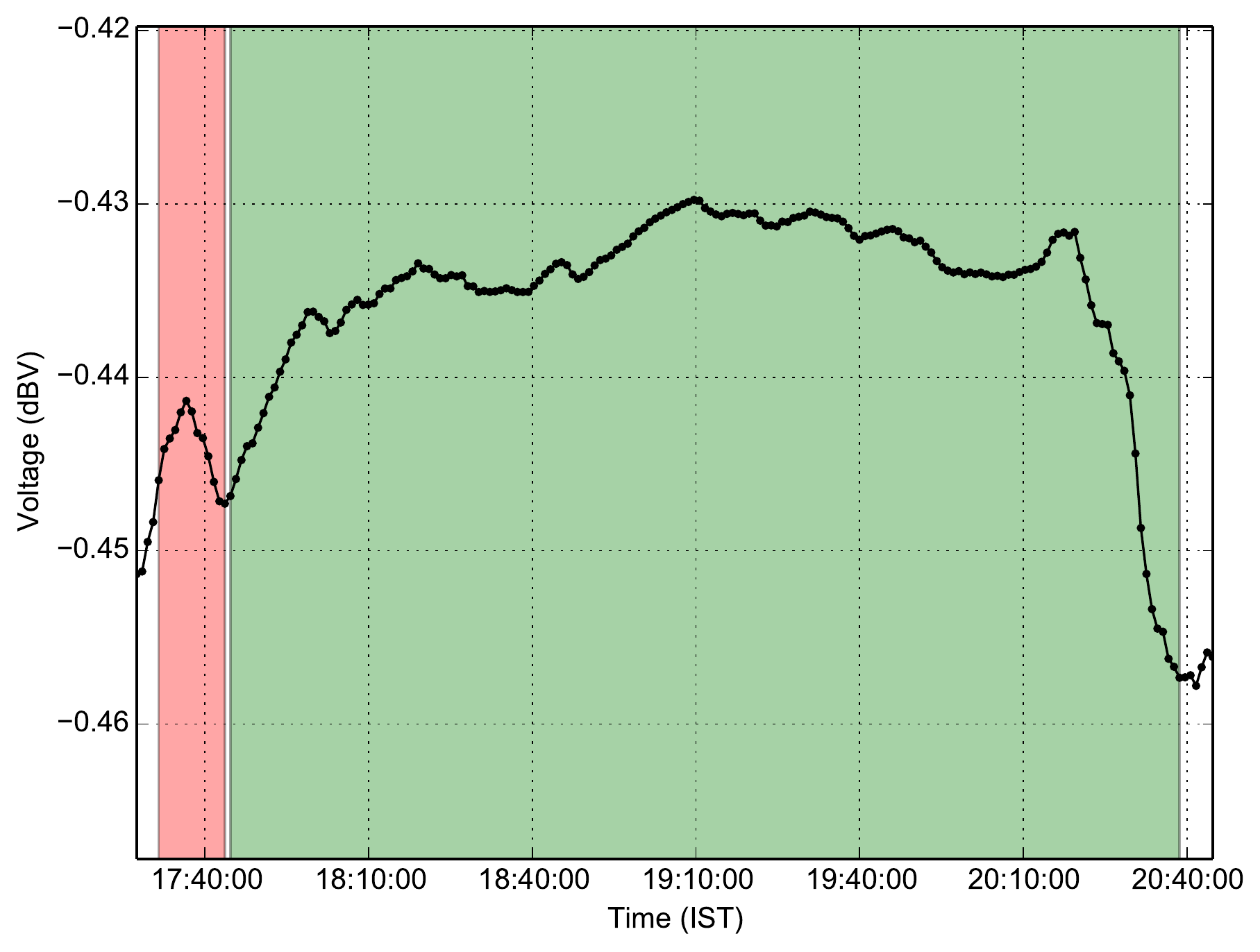}
\caption{Drift scans Saturn and Milky way. The pink and the green region corresponds to Saturn and the Galactic arm respectively.}
\label{satgalplot}
\end{figure}

The radiometer equation gives us 
\begin{equation}
\label{radiometer}
\sigma_{T}\propto\frac{1}{\sqrt{\tau \Delta \nu_{RF}}}
\end{equation}
Where, $ \sigma_{T} $ is RMS receiver output fluctuation, $ \Delta \nu_{RF} $ is the bandwidth and
$ \tau $ is the integration time.\cite{editorsite}
The equation tells the reduction in noise when integration time is increased thus giving us a higher sensitivity. We increased our integration time from 1 second to 60 seconds, and drift scans were taken and for Saturn and galactic arm (as shown in Fig. \ref{satgalplot}). We report the detection of Saturn with an signal to noise ratio (SNR) of 3. The sources were confirmed by correlating the time of arrival of the source signal in the field of view of our telescope with the predicted times by Stellarium based on our pointing.\cite{stellarium} Fig.\ref{stellarium} shows the approximate path traced by the field of view of the telescope during the drift scan observation with respect to Fig \ref{satgalplot}.
%%%%%%%%%%%%%%%%%%%%%%%%%%%%%%%%%%%%%%%%%%%%%%%%%%%%%%%%%%%%%%%%%%%%%%%%%%%%%%%%%%%%%%%%%%%%%%%%%%%%%%%%%%%%%%%%%%%%%%%%%%%%%
\section{Discussion}
This radio telescope requiring only everyday items that make it affordable and a versatile tool to demonstrate the working of a radio telescope. The methods and calculations shown here give an all-round hands-on experience about fundamentals of radio telescopes, computer programming and data analysis with an inherent amateur astronomy in it. This radio telescope shows good response when customizations are made, which provides an opportunity to study the properties of the telescope itself. It has a very good response to the signal from geostationary satellites, and the pointing towards such point sources are easier due to the large beam width.\\

Sun, being a strong and easily detectable radio source, is easy to point at and study. Drift-scan for the sun was performed, and data was fitted with a Gaussian. This reveals an important property: the antenna beam width and provides a coherent idea about it. When backed up with satellite observation (preferably near the same time) the same drift-scan can be used to measure source flux and brightness temperature.\\

The radiometer equation (equation \ref{radiometer}), shows that the noise is inversely proportional to the square root of the bandwidth of the telescope and the integration time. Thus by increasing the integration time, we can reduce noise. Most popular ammature telescopes like VSRT use RTL-SDR, which has a bandwidth of $\sim$ 3 MHz\cite{rtl}, whereas, with the satellite finder we can utilize the full 1.1GHz bandwidth, thus giving us a higher sensitivity. Although most undergrad institutions have such amateur astronomy programs, but radio astronomy is not popular among them. As we have shown, increase in sensitivity would give one the opportunity to look at many other sources than just the Sun. All-sky visibility maps can be generated at a rudimentary level that would make one familiar with a bigger part of the radio sky.\\
\begin{figure}[!h]
\centering
\includegraphics[scale=0.65]{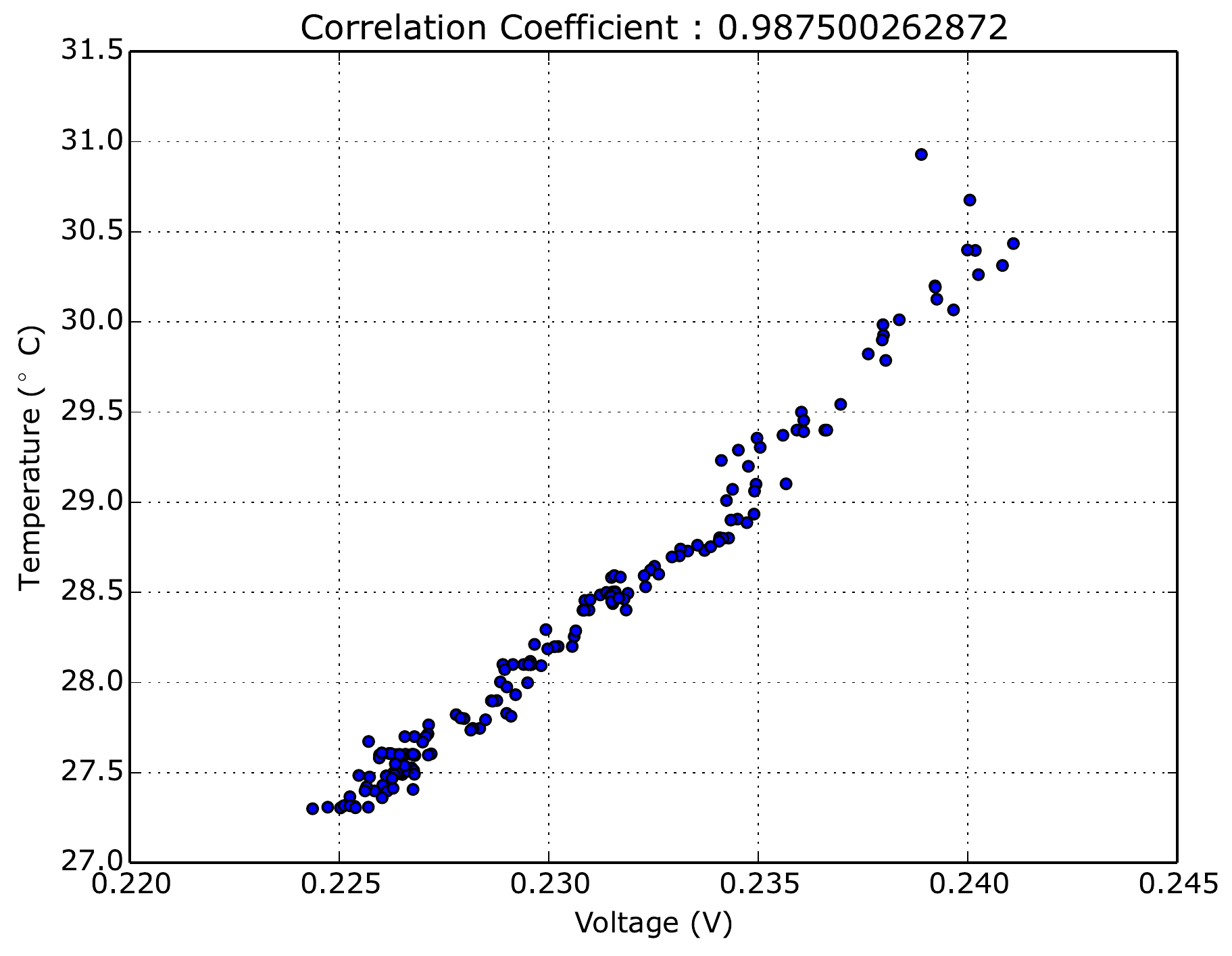}
\caption{Correlation between surrounding temperature and voltage.}
\label{corr}
\end{figure}
We measured the temperature of the LNB surroundings during observations. An increase in the voltage was observed with the rise in temperature. The voltage and the surrounding temperature plot increases linearly as shown in \autoref{corr}. The correlation coefficient between these two was obtainted to be 0.98, confirming the linear dependence of voltage on the surrounding temperature change.
%%%%%%%%%%%%%%%%%%%%%%%%%%%%%%%%%%%%%%%%%%%%%%%%%%%%%%%%%%%%%%%%%%%%%%%%%%%%%%%%%%%%%%%%%%%%%%%%%%%%%%%%%%%%%%%%%%%%%%%%%%%%%
\begin{acknowledgments}
This work is supported by School of Physics IISER-Thiruvananthapuram. The authors would like to thank Joy Mitra and N. Rathnasree for useful discussion and helpful comments on this work. We would also like to thank S. Shankarnarayanan and M. Thalakulam for their support. AY, KS, PK, and PM are supported by INSPIRE fellowship. DA thanks the Max Planck Partner Group Research Facility at IISER-Thiruvananthapuram. A part of the manuscript was written while being a Project Assistant at the MPPG on GWaves at IISER-Thiruvananthapuram.
\end{acknowledgments}
%%%%%%%%%%%%%%%%%%%%%%%%%%%%%%%%%%%%%%%%%%%%%%%%%%%%%%%%%%%%%%%%%%%%%%%%%%%%%%%%%%%%%%%%%%%%%%%%%%%%%%%%%%%%%%%%%%%%%%%%%%%%%

\appendix
%%%%%%%%%%%%%%%%%%%%%%%%%%%%%%%%%%%%%%%%%%%%%%%%%%%%%%%%%%%%%%%%%%%%%%%%%%%%%%%%%%%%%%%%%%%%%%%%%%%%%%%%%%%%%%%%%%%%%%%%%%%%%
\section{Appendix : Temperature Calculation}
\label{appendix1}
We provide the detailed temperature calculation here. The Friis transmission equation gives a relation between power received by antenna, $ P_{r}$ of gain $ G_{r}$ when an antenna of gain $ G_{t}$ is transmitting a power $ P_{t} $.
\begin{eqnarray}
\dfrac{P_{r}}{P_{t}}= G_{t} G_{r} \left(\dfrac{\lambda}{4 \pi R}\right)^{2}
\end{eqnarray}
Here, $ \lambda$ is the observing wavelength and $R$ is the distance between the transmitter and the receiver antenna. For our system $\lambda = 2.72 \times  10^{-2}$ m. Distance to the satellite and EIRP values in the region are freely available on various websites.\cite{dishpointer}
\begin{eqnarray}
P_{r}&=&P_{t} G_{t} G_{r} \left(\dfrac{\lambda}{4 \pi R}\right)^{2} \nonumber \\
\text{EIRP}&=&P_{t} G_{t} \qquad(\text{for satellite} ) \nonumber  \\
P_{r}&=& \text{EIRP}\, G_{r} \left(\dfrac{\lambda}{4 \pi R}\right)^{2} 
\end{eqnarray}
For INSAT-4A, EIRP = 52 dBW in Trivandrum, India and R = 3.62657 $\times 10^{7}$ m. Gain for an antenna is defined as.
\begin{eqnarray}
G_{r}=\eta_{A} \left(\dfrac{\pi d}{\lambda}\right)^{2}
\end{eqnarray}
where $\eta_{A}$ is the aperture efficiency which is the ratio of the antenna's effective aperture $A_{e}$ to the antenna's physical aperture area $A_{p}$:
$\eta_{A} = \frac {A_{e}} {A_{p}} $ and``d'' is the diameter of the dish. For our setup $\eta_{A}=0.55$.%(as given with antenna when it was bought).
\begin{eqnarray}
G_{r}&=& 0.55 \times \left(\dfrac{\pi \times 0.71}{2.72 \times 10^{-2}} \right)^{2} = 3698.63
\end{eqnarray}
For INSAT-4A
\begin{eqnarray}
P_{r}&=&EIRP \times  G_{r}  \times \left(\dfrac{\lambda}{4 \pi R}\right)^{2} \nonumber \\
&=&10^{5.2} \times 3698.63 \left(\dfrac{2.72 \times 10^{-2}}{4 \pi \times 3.62 \times 10^{7}}\right)^{2} \nonumber \\
&=& 2.0957 \times 10^{-12 } W
\end{eqnarray}
While pointing at a source, we measure the increase in voltage with respect to cold sky. We now standardize the power received for a unit increase in voltage for a bandwidth of 400MHz as the satellite emits only for this bandwidth and assume this extrapolates linearly for whole 1.1 GHz bandwidth. Here $\Delta V_{sat}=V_{sat}-V_{bground}$ and $\Delta V_{sun}=V_{sun}-V_{bground}$
\begin{eqnarray}
1V&=&\dfrac{P_{r}}{\Delta V_{\text{sat}}}\\
\Delta V_{\text{sat}}&=& 0.0980 V \quad \Rightarrow 1V = 2.1384 \times 10^{-11}W
\end{eqnarray}
Power received from the sun is now calculated :
\begin{equation}
\Delta V_{\text{sun}}= 0.0521 V \quad \Rightarrow P_{t} = 11.1410\times 10^{-13} W
\end{equation}
Power density, $P_d$ and flux calculations are as follows.
\begin{eqnarray}
\text{where}\; A_{e}&=&\frac{G_{r}\lambda^{2}}{4 \pi} = 0.2177 m^{2},\, \text{and}\; P_{d}= \frac{P_{t}}{A_{e}} = 5.1175 \times 10^{-12} W/m^{2}\\
\text{flux}&=&\frac{2 \times P_{d} \times 2.75}{ \Delta \nu_{RF} } = 2.5587 \times 10^{6} Jy
\end{eqnarray}
where $ \Delta \nu_{RF} $=1.1 GHz\\
We include a factor of 2 because we only observe one polarization. INSAT-4A emits in 400 MHz band while emission from the sun is broadband in nature emitting for whole 1.1GHz bandwidth. Therefore, we include a factor of 2.75 for solar flux calculation.
We assumed that the Sun's radiation is thermal and used Rayleigh -Jeans law to calculate the temperature. The intensity, $I$
\begin{eqnarray*}
I = \dfrac{Flux}{\Omega} =3.7599\times 10^{10} Jy/Sr
\end{eqnarray*}
where ${\Omega}$ = 6.8052$\times 10^{-5}$ Sr , is the solid angle subtended by the sun.\\
and the temperature, $T$
\begin{eqnarray*}
T&=&\dfrac{I\lambda^{2}}{2K_{B}} = 10078\,K
\end{eqnarray*}

%\subsection{Error calculation}
%\begin{eqnarray}
%\Delta V_{sat/sun}&=&V_{on source}-V_{off \,source}\\
%\delta (\Delta V)&=&\sqrt{{(\delta V_{on source})}^{2} + {(\delta V_{off \,source})}^{2}}\\
%\delta (1 V)&=&\dfrac{P_{r} \times (\delta(\Delta V_{sat}))}{\vline\ {\Delta V_{sat}}\vline\ \times 2.75}
%\end{eqnarray}
%This factor of 2.75 comes because $P_{r}$ as well $\Delta V_{sat}$ both increase by a factor of 2.75 as both are corresponding to 400MHz bandwidth instead of 1.1GHz.
%\begin{eqnarray}
%\delta P_{t}&=&\sqrt{{(\dfrac{\delta (1 V)}{1 V}})^{2} + (\dfrac{\delta(\Delta V_{sun})}{\Delta %V_{sun}})^{2}}*P_{t}\\
%\delta T&=& (T \times \delta P_{t})/P_{t}
%\end{eqnarray}
\begin{figure}[h]
\centering
\includegraphics[width=1.3\textwidth, angle=90]{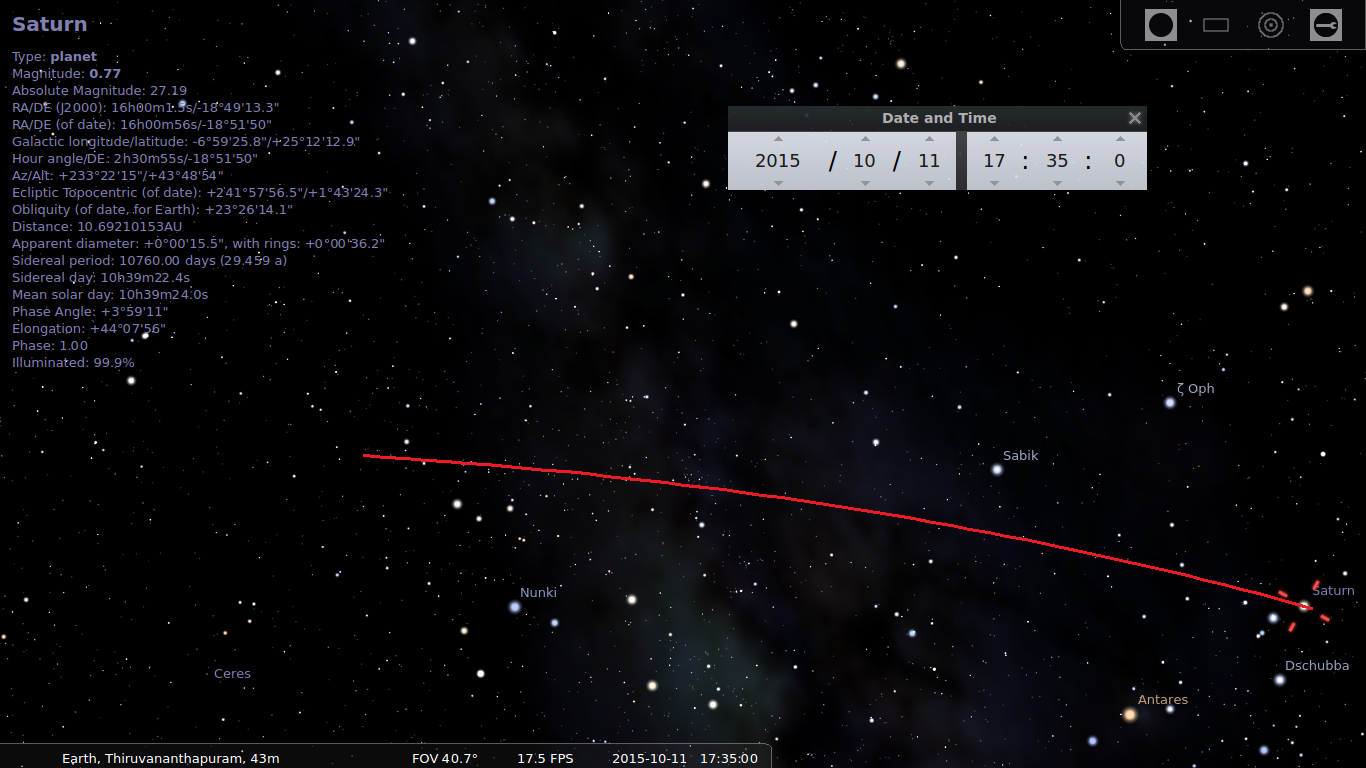}
\caption{The red line show the approximate path traced by the field of view of the telescope during the observation shown in Figure \ref{satgalplot}. The coordinate of Saturn at the time of it's passing is shown on top left corner.}
\label{stellarium}
\end{figure}
\end{document}